\documentclass[twocolumn,a4paper,10pt]{article}

\usepackage[english]{babel}
\usepackage{fullpage,amsmath,amssymb,amsfonts,epsfig, subfigure, float, setspace}
\renewcommand\thefootnote{\fnsymbol{footnote}}
\numberwithin{equation}{section}
\psfigdriver{dvips}

\begin{document}
\title{Mutual information in the Tangled Nature Model}

\author{Dominic Jones,  Henrik Jeldtoft Jensen\footnotemark[1] \\
{\small Institute of Mathematical Sciences, Imperial College London}\\
{\small London SW7 2PG, UK}\\
{\small and}\\
{\small Department of Mathematics, Imperial College London}\\ 
{\small London SW7 2AZ, UK}\\
Paolo Sibani \\ 
\small{Institut for Fysik og Kemi, SDU, DK5230 Odense M, Denmark}}
\date{\today}

\twocolumn[
\begin{@twocolumnfalse}

\maketitle

\begin{abstract}
We consider the concept of mutual information in ecological networks, and use this idea to analyse the Tangled Nature model\ of co-evolution. We show that this measure of correlation has two distinct behaviours depending on how we define the network in question: if we consider only the network of viable species this measure increases, whereas for the whole system it decreases. It is suggested that these are complimentary behaviours that show how ecosystems can become both more stable and better adapted. 
\end{abstract}
\end{@twocolumnfalse}
]

\fnsymbol{footnote}
\footnotetext[1]{Author for correspondence (h.jensen@imperial.ac.uk)}
\renewcommand\thefootnote{\arabic{footnote}}

\section{Motivation}  

Identifying universal features of ecosystem dynamics has been a long-standing goal in ecology. These attempts have usually involved identifiying system variables that are potentiallly optimised during the evolution of an ecosystem. Many such candidate variables have been identified. Increasingly the focus has been on the network properties of the ecosystem, or more precisely the trophic net defined by the mass flows between the species constituting the ecosystem. However empirical evidence at the resolution needed to verify any particular claim remains out of reach for most studies. For ecologists these quantities are both of theoretical and practical interest. From a theoretical point of view it would be nice,as already noted, to find some governing principle of ecological dynamics, while practically speaking there is a need to establish a good measure of ecosystem health and maturity \cite{Ulanowicz2002, Christensen1995}.

In this paper we propose to study this issue in the context of a well established evolutionary model. The Tangled Nature model of co-evolution\cite{Christ2002} has already been studied in several contexts \cite{Hall2002,Jensen2005,Jensen2006} and is ideal for this work as it is designed specifically to study long time behaviour in ecological networks. Its simplicity along with the rich complexity of its resulting behaviour makes it a paradigmatic model for testing co-evolutionary ideas. The model retains  the binary string genotype geometry found in previous 
approaches (for example the quasispecies model \cite{Eigen1977}or the NK model \cite{Kauffman1990}, but replaces their `ad hoc'  static fitness  landscapes with a set of population  dependent interactions between extant species, similar to  the `tangled' interactions of an eco-system. From a `random' initial state,  the network of extant and interacting
population  changes over time, slowly, but radically, enabling the system to support an ever growing number of individuals.

Despite its simplicity, the model is able to reproduce the long time decrease reported in the overall macroscopic extinction rate, the observed intermittent nature of macro-evolution, denoted punctuated equilibrium by Gould and Eldredge, the log-normal shape often observed for the Species Abundance Distributions, the power law relation often seen between area and the number of different species number, the framework of the model is also able to reproduce often reported exponential degree distributions of the network of species as well as the decreasing connectance with increasing species diversity that has attracted much observational and theoretical interest.      

The details of the model are described in greater detail below, but the key aspect of its behaviour is that it moves through a series of different network configurations. In this paper we analyse these dynamic networks using tools developed in ecology. In particular, we are able to shed light on the tension between robustness and efficiency in ecological networks highlighted by Jorgensen et al~\cite{Jorgensen2007}. Increased correlation lead to greater brittleness in the case of perturbations, but greater robustness leads to an apparent squandering of resources. We suggest how this conflict can be resolved using evidence from Tangled Nature, where it is possible to divide the system into two interacting parts - a viable network of keystone species, and a periphery of unviable mutants. Seen from this perspective the apparent paradox is resolved, as the viable network becomes increasingly correlated, while the total network (including many species \emph{in potentia}) develops greater redundancy.

\section{Review of the basic behaviour of the model}
\subsection{Type space and the interaction matrix}
A type is represented by a vector $S$ of $L$ elements belonging to the set $[0,1]$. Thus there are $2^L$ possible types, corresponding to the vertices of a unit hypercube in L-dimensions. $S$ may be interpreted as a genome, or a set of characteristics - either way it is directly susceptible to mutations and defines the type completely (that is there is no phenotype level in this model). Each type, which we can index by a number $i$ in the range $1 - 2^L$ to simplify notation, has a population of $n_i(t)$ identical individuals, so the total population is the sum over all the $2^L$ possible types 

\begin{equation}
{
N(t)=\sum_{i=1}^{2^L}n_i(t)
}
\end{equation}
The ability of an individual to reproduce is determined by how it interacts with the other types present at a given time. This is formalised in the reproduction weight function (which is then turned into a probability of reproducing - see below)
\begin{equation}
H_i(t)  =  \frac{c}{N(t)}\sum_{i=1}^{2^L} J(S_i, S)n(S_i,t) - \mu N
\end{equation}
where the sum is over all other types, $C$ is a control  parameter that determines the level of inhomogeneity in the population, $N(t)$ is the total population at time $t$, and $n(S,t)$ is the population of type $S$.

\begin{figure}[htbp]
\centering
\includegraphics[width=8cm]{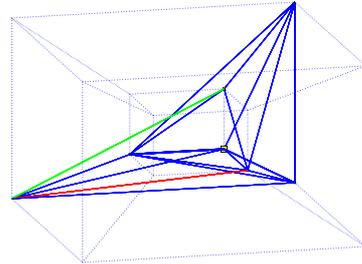}
\caption{An example of the configuration of the Tangled Nature system in a meta-stable state. This is a 4 dimensional model for expository purposes only, the model in this paper has 20 dimensions. The vertices of the hypercube represent the 16 possible types in the model. The dotted lines represent nearest neighbour links in type space, and the solid lines represent non-zero interaction terms with blue = +-, red = --, green = ++}\label{fig:egintmat}
\end{figure}

Two types $S_i$ and $S_j$ are coupled via the interaction matrix $J(S_i, S_j)$ that can be either positive negative or zero. This number is intended to be the sum of all the influences of $i$ upon $j$. This interaction matrix is unrelated to the type space outlined above so there are no correlations in the interactions between different types - that is $<J_(S_i, S_j)J(S_k,S_j)> = 0$ even if the average is restricted to neighbours in type space. This interaction is not necessarily material in nature but may represent any influence that one type has on another. The overall connectivity of the interaction matrix is set by a parameter $\Theta$ which for this paper has a value of 0.2 (that is 0.2 of all possible connections between types actually exist). The distribution of the nonzero values of the function $J(S_i,S_j)$ are irrelevant as long as they are distributed in some reasonable, continuous way. 
The interaction matrix is constructed such that if $J(S_i, S_j)$ is nonzero then $J(S_j,S_i)$ is also nonzero. This means there are three types of interaction - mutualistic, antagonistic and predator-prey.
Figure~\ref{fig:egintmat} illustrates the key components of the tangled nature model - the hypercubic type space, varying type occupancies, and the different types of possible interaction between types.

\subsection{Reproduction, mutations and death}
The model is simulated stochastically, with a time-step consisting of the following: one individual is selected at random, and reproduces asexually acccording to the probability

\begin{equation}
P_r(S_i,t)=\frac{1}{1+exp[H(S_i,t)]} \in [0,1] 
\end{equation}

If successful the individual is replaced with two copies. In each of these copies there is a probability of mutation per `gene', $p_m$. Another individual is picked at random and is killed with probability $p_k$. 

\subsection{General behaviour of the model}
\begin{figure}[htbp]
\centering
\includegraphics[width=8cm]{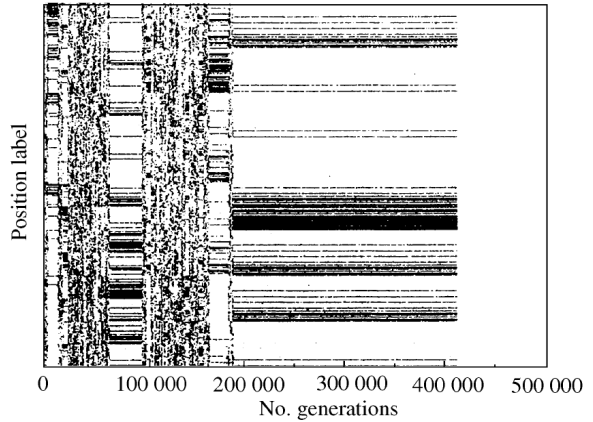}
\caption{Overview of a typical run of TaNa. The y-axis is simply a species label, ranging from 1 - $2^L$, and the $x$-axis is time in generations. If a position is occupied at a given time, a dot is placed at the corresponding number for that time step. The plot clearly shows the alternating stable and unstable periods. The stable periods are characteristised by a steady population and constant set of species, whereas the transitions have a constantly changing set of species (eg between 100 000 and 150 000 generations)  Figure from \cite{Christ2002}. }\label{fig:specdist}
\end{figure}

We start a run with $N(0)=1000$ individuals on one randomly chosen site. Initially there is no reproduction, since there can be no interactions between species, so $H$ is very negative and the probability of reproduction is zero. Then as the resource limitation term diminishes, reproduction becomes possible, and consequently some new types are generated by mutations. Once interactions between these new types begins, the interaction term in the reproduction probability becomes significant. After some re-organisation, a set of species that interact in a stable way emerges, and persists for some time (see figure~\ref{fig:specdist}). This period of stability is ended by another chaotic reorganisation, from which another meta-stable state emerges.  

\begin{figure}[htbp]
\centering
\includegraphics[width=8cm]{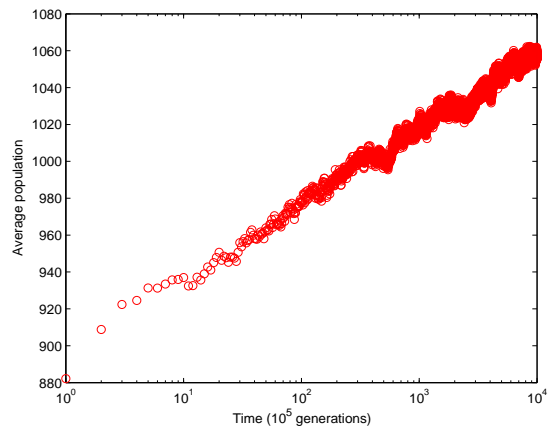}
\caption{The mean population (averaged over an ensemble of 1000 runs) increases logarithmically in time}\label{fig:avepop}
\end{figure}

The bulk properties of these meta-stable states turn out to depend on the age of the system - the system slowly optimises the interactions between species, as evidenced for example by the logarithmically increasing population (figure~\ref{fig:avepop}). It is this non-stationary aspect of the model that this paper tries to explain, albeit only partially.

\section{Results}
We ran 1500 simulations of the model with an initial population confined to one randomly chosen site. The random interaction matrix was regenerated each time. The parameters used for all the runs were the same, and were chosen to robustly generate the intermittent regime for a population of a manageable size. 

We use the following parameter values: $ \mu = 0.14,~p_{mut}=0.03,~p_{kill}=0.2,~c=10$. Detailed discussion of the various regimes defined by these parameters can be found elsewhere; for now we simply note that the behaviour generated by this set is characteristic of a significant area of parameter space. The one major change is seen when $p_{mut}$ goes above the error threshold, which results in diffusion dominated behaviour.

\subsection{The Core and the Periphery}

\begin{figure}[htbp]
\centering
\includegraphics[width=8cm]{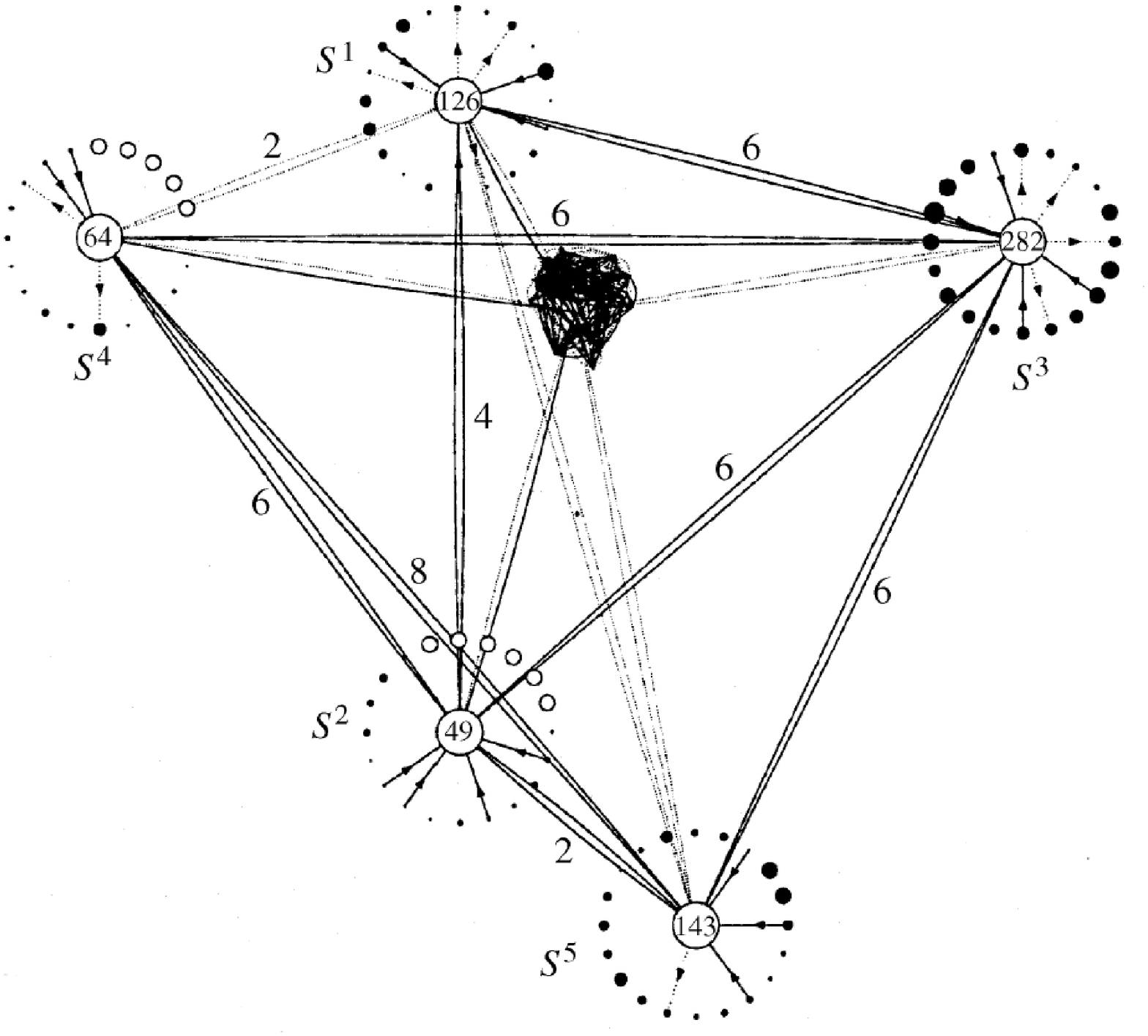}
\caption{A typical stable ecological network seen in the Tangled Nature model. Each circle represents a species with the size proportional to the current population. Lines represent interaction links ($|J|>0$) between species. The mutation network has been suppressed to clarify the figure but the line indices give the hamming distance (a measure of evolutionary separation) between each species. The core is the network of large nodes, and the whole system  includes the smaller unlabelled nodes. Figure from \cite{Christ2002}.}\label{percore}
\end{figure}

The network realised at any given time can be divided into two classes - those nodes that are viable (loosely, those that have a birthrate approximately equal to the death rate) and those that aren't. This second group are the mutants from the viable core, who in the current configuration are not able to reproduce. Figure~\ref{percore} schematically depicts this arrangement, with each viable species having a flower of unviable mutants surrounding it. These mutants do not, in general play an active role (even as a stabilising factor) during a stable period, but they are in the end responsible for the eventual collapse of one metastable state and creation of another. The following results are obtained for both the whole system, and the viable core.

\subsection{Mutual Information}

Ideas from information theory have  been used in ecology for over 50 years \cite{Macarthur1955} \cite{Wilhelm2007}, and Rutledge et al introduced the idea of using the mutual information of networks as a measure of their stability \cite{Rutledge1976}. This was all somewhat unnoticed by those working more recently on networks in graph theory and complexity. This is principally due to the fact that ecologists must work with weighted networks, whereas  most recent work on network characterisation has focussed on unweighted networks, for which there exist a large arsenal of analytical tools.

First we define what the mutual information is for a general random process, then we will define how we use this measure in this paper. The information of a realisation $x$ of a random variable $X$ is defined via its probability distribution $P(x)$, as 
\begin{equation}
I(x)=P(x)\text{log}P(x)
\end{equation}
For two random variables, we can define the mutual information, which is defined as the reduction in the uncertainty of $X$ given knowledge of $Y$. The mutual information is defined on two random variables $X$ and $Y$ as
\begin{equation}
I(X,Y)=\displaystyle \sum_{x,y}P(x,y)\log  \left(\frac{P(x,y)}{P_1(x)P_2(y)}\right)
\end{equation}
 where $P_1$ and $P_2$ are the marginal distributions of $X$ and $Y$ respectively, and $P$ the joint probability distribution. Equally we can think of the mutual information as the constraint imposed on $X$ by $Y$. 

The Tangled Nature model is a model of network evolution. As the structure of the network changes, we ask the question: how does the current network structure constrain its evolution? The network we consider is the interaction network J weighted by the occupancy of the species, so that we only consider connections between extant species. When this condition is met, we consider there to be $n_in_j$ copies of link $J_{ij}$. Consider the ensemble link value distribution at time $t$, $P(J,t)$. This gives the probability of a link value $J$ for an ensemble of realisations. However if we consider a particular realisation, we can expect that this distribution, $P(J,t,r)$ (where r indexes specific realisations) will in general differ from the ensemble average. We can measure this difference by looking at the joint probability distribution $P(J_1,J_2,t,r)$. The degree to which this quantity differs from the product of the marginal distributions for $J_1$ and $J_2$ (which in our case are identical, equal to the distribution over the ensemble $P(J,t)$) measures the degree to which the presence of some link value $J_1$ influences the presence of some other value $J_2$.

\begin{figure}[htbp]
\centering
\subfigure[The mutual information of the whole network as a function of time. We see a slow but significant decrease signifying a decorrelation of the component parts of the network.]{
\includegraphics[width=7cm]{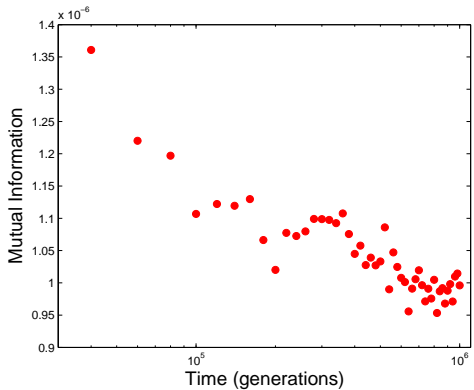}
\label{fig:MI1}
}
\subfigure[The mutual information of the core  as a function of time. For this subset the mutual information increases over time, indicating greater correlation and efficiency.]{\includegraphics[width=7cm]{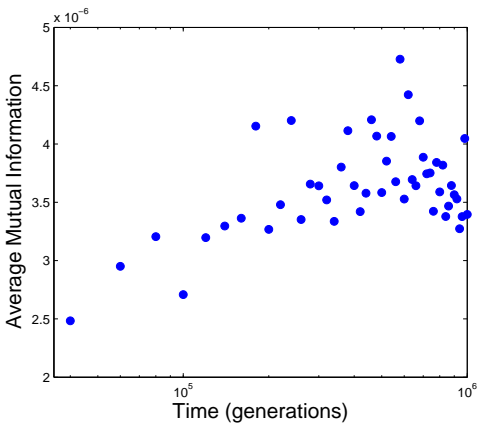}
\label{fig:MI2}
}
\subfigure[The mutual information of a random set of networks using the same average diversity and population as the simulation results. The mutual information has no trend and is approximately 2 orders of magnitude smaller.]{\includegraphics[width=7cm]{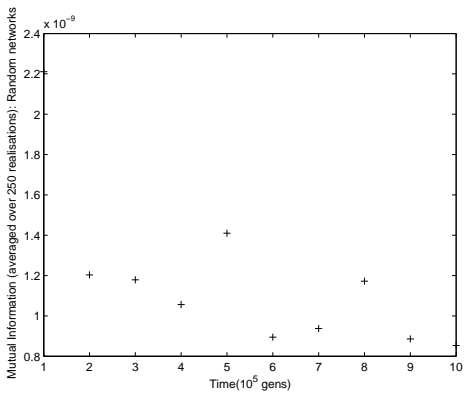}
\label{fig:MIR}
}
\caption{Contrasting trends in the mutual information for different subsets}
\end{figure}
To consider the probability of a link value appearing at time $t$, we first introduce a new variable which will simplify the following. We will consider a single index $k$ that runs over all links in a realisation, and each link is waited by $d_k$, the product of the occupancy of the two nodes at either end: $d_k = d(J_{ij})=n_in_j$.
Explicitly we define the relevant quantities as follows: the probability that the link value $J$ appears at time $t$ is

\begin{equation}
P(J,t) = \frac{1}{DR}\sum_{k,r} d_k\delta(J-J_k) 
\end{equation}

where $D$ is the number of links counted between all extant species, $D=\sum_{i,j} n_in_j$
and $R$ is the number of realisations.
 whereas the joint probability distribution for two link values to appear in one realisation is

\begin{equation}
 P(J_1,J_2,t,r)=\frac{1}{D}\sum_{k,l} (d_k + d_l) \delta(J_1-J_k) \delta(J_2-J_l)
\end{equation}

 With these quantities defined, we may define the mutual information on these distributions as
\begin{equation}
I(J,J',t)=\displaystyle \sum_{r,i,j} P(J_i,J_j,t)\log \left(\frac{P(J_i,J_j,t,r)}{P(J_i,t)P(J_j,t)} \right)
\end{equation}
Since the link distribution fluctuates due to the stochastic nature of the system, this distribution is calculated over a small time window $\delta t $ where $\frac{\delta t}{t_{\text{max}}}<<1 $.

Figures \ref{fig:MI1} and   \ref{fig:MI2} show the evolution of the mutual information over time for two different subsets of the system. Figure \ref{fig:MI1} is the MI for the whole system, where we see a declining trend. The subset of vertices linking nodes with more than 5 individuals by contrast displays an increase in the MI over time (figure~\ref{fig:MI2}). 

We note that in general the mutual information is quite low, which is expected. We are measuring the influence of the presence of link values on the presence of other link values; this influence is highly constrained by the quenched randomness of the network and the stochastic dynamics, so in general we do not expect the mutual information to be high. Nevertheless we have compared the values obtained to simulated random networks of equivalent size and connectance, and found the mutual information to be approximately three orders of magnitude smaller. 

The data is significantly noisy despite being the result of a large ensemble average. Nevertheless it is clear, especially for the whole system, that the curves are approximately linear in logarithmic time. This corresponds to the behaviour of other measures of the system, and can possibly ultimately be related back to some record process.

Averaging over more realisations increased the clarity of the results, but at the cost of computing time. To decrease the fluctuations by an order of magnitude would have required approximately 400 weeks more computing time.

\section{Discussion}  
The question of how the structure of an ecosystem, or any system of interacting, evolving agents, changes over time is a controversial one, and to some extent depends on the details of the system under consideration. In this paper we have considered a generic evolutionary model with the aim of elucidating ecological dynamics in the general case. The apparent competition between two requirements of a viable ecosystem - that they maximise resource us on the one hand, and remain robust to perturbations on the other - poses the question: what in fact happens? 

The obvious way to answer this question would be to do an experiment. However, ecological experiments of the type required (both in terms of detail and time resolution) are not currently possible. Indeed, ecological data recorded over evolutionarily significant timescales is practically unattainable for any but the fastest evolving systems, such as microbial populations (see for example \cite{Lenski2008}). However, even for such experimentally manipulable systems it may be hard to infer interaction networks accurately. The practical difficulties of experiments in evolutionary ecology is one of the key reasons why we believe theoretical work such is that presented is important, since it can act both as spur and guide for future experimental work.

We have found that while the ecosystem as a whole becomes less correlated over time, the correlation of the network of its core species increases. While we have not shown it here, it seem plausible that this is two sides of the same coin - decorrelation of the whole system implies that the system explores a greater range of possible networks, from which it chooses more and more well correlated subsets. This fits with other results we have obtained that show the model increases its population over time. 

When considering ecological networks, most work has naturally focussed on trophic networks, that is networks of material flow through an ecosystem. This has yielded a natural way to analyse these networks, since the dynamics is conservative, one can consider the probability of any two species being involved in material exchange. The Tangled Nature model explicitly models more than simply mass flow in ecosystems: it attempts to quantify the \emph{influence} that one species has on another. While this has the advantage of allowing one to consider more than simply predator-prey relationships (for example mutualistic behaviour arises very naturally in the model), it means that one cannot simply take over tools used on trophic nets wholesale. In this paper we have adapted the approach used in ecology and elsewhere to this interaction view with the caveat that our results are not directly comparable to those gleaned from analysis of food webs; we did also attempt to interpret the model as a flow model but found that this approach yielded no clear information about the network structure. One possibility in this direction is to adopt the approach in \cite{Demetrius2007} where once a network has evolved one imagines some simple Markovian dynamics entirely independent of the actual model dynamics in order to determine the relevant network measures. 

We have not used any of the more simple information theoretic measures available ( for example the entropy). This is because we found it necessary to consider a quantity that characterised the difference of a specific realisation from an ensemble of realisations. The entropy of the system as a whole increases over time, but there is no corresponding decrease in the core population. It is easy to see why: the entropy over an ensemble of realisations is simply the sum of individual realisations and so one would only expect to see a decrease in entropy if \emph{every} realisation converged on a small set of link values. This by no means has to be the case, since the system can adjust species populations to a wide range of networks. The mutual information, on the other hand, measures how the existence of certain links within one realisation determines the presence of other links \emph{within that same realisation} and so does increase over time. It remains to be seen whether there is some entropic measure in Tangled Nature (or indeed in reality) which is maximised through evolution.

One might naively think that the result for the core is simply due to the increasing stability of the system observed in other contexts. Taken by itself this is reasonable, since it is possible that the system stabilises over time, and that this stabilisation would positively contribute to mutual information of the core. However, if it was purely an artefact of the system spending more time in a stable configuration then we would expect the whole system (that is both the viable core \emph{and} the surrounding mutants in figure ~\ref{percore}) to display a similar positive trend, which is clearly not the case. Therefore we conclude that the increasing correlation of the core, along with the increasing decorrelation of the periphery of the system, plays a causal role in the stabilisation of the system as a whole. We postulate that these two phenomena are linked - the system explores a greater number of possible links which allows it to find better adapted sets of links for the core, which in turn leads to a bigger population and an even larger set of links to select from. While we do not claim to have proved that this is the case, the data is strong evidence that some adaptive behaviour of this type is occurring. In future papers we hope to probe the nature of this adaptive dynamics further.


\begin{thebibliography}{10}

\bibitem{Ulanowicz2002}
Robert~E. Ulanowicz.
\newblock The balance between adaptability and adaptation.
\newblock {\em Biosystems}, 64(1-3):13 -- 22, 2002.

\bibitem{Christensen1995}
Villy Christensen.
\newblock Ecosystem maturity - towards quantification.
\newblock {\em Ecological Modelling}, 77(1):3 -- 32, 1995.

\bibitem{Christ2002}
Kim Christensen, Simone A.~Di Collobiano, Matt Hall, and Henrik~J. Jensen.
\newblock Tangled nature: A model of evolutionary ecology.
\newblock {\em Journal of Theoretical Biology}, 216(1):73 -- 84, 2002.

\bibitem{Hall2002}
Matt Hall, Kim Christensen, Simone~A. di~Collobiano, and Henrik J.~Jensen.
\newblock Time-dependent extinction rate and species abundance in a
  tangled-nature model of biological evolution.
\newblock {\em Phys. Rev. E}, 66(1):011904, Jul 2002.

\bibitem{Jensen2005}
Paolo Sibani and Henrik~J. Jensen.
\newblock Intermittency, aging and extremal fluctuations.
\newblock {\em EPL (Europhysics Letters)}, 69(4):563--569, 2005.

\bibitem{Jensen2006}
Daniel Lawson and Henrik~J. Jensen.
\newblock The species-area relationship and evolution.
\newblock {\em Journal of Theoretical Biology}, 241(3):590 -- 600, 2006.

\bibitem{Eigen1977}
Manfred Eigen and Peter~K. Schuster.
\newblock A principle of natural self-organisation.
\newblock {\em Naturwissenschaften}, 64:541--565, 1977.

\bibitem{Kauffman1990}
Stuart~A. Kauffman.
\newblock {\em The Origins of Order: Self Organisation and Selection in
  Evolution}.
\newblock Oxford University Press, 1990.

\bibitem{Jorgensen2007}
Sven~E. Jørgensen, Simone Bastianoni, Brian~D. Fath, Felix Muller, Joao~C.
  Marcques, Soren~N. Nielsen, Bernard~C. Patten, Enzo Tiezzi, and Robert~E.
  Ulanowicz.
\newblock {\em A New Ecology: Systems perspective}.
\newblock Elsevier, 2007.

\bibitem{Macarthur1955}
Robert MacArthur.
\newblock Fluctuations of animal populations and a measure of community
  stability.
\newblock {\em Ecology}, 36(3):533--536, 1955.

\bibitem{Wilhelm2007}
Thomas Wilhelm and Jens Hollunder.
\newblock Information theoretic description of networks.
\newblock {\em Physica A: Statistical Mechanics and its Applications},
  385(1):385 -- 396, 2007.

\bibitem{Rutledge1976}
Robert~W. Routledge, Bennett~L. Basore, and Robert~J. Mulholland.
\newblock Ecological stability: An information theory viewpoint.
\newblock {\em J. Theor. Biol.}, 57:355--371, 1976.

\bibitem{Lenski2008}
Zachary~D. Blount, Christina~Z. Borland, and Richard~E. Lenski.
\newblock Historical contingency and the evolution of a key innovation in an
  experimental population of escherichia coli.
\newblock {\em PNAS}, 105(23):7899--7906, 2008.

\bibitem{Demetrius2007}
Lloyd Demetrius and Martin Ziehe.
\newblock Darwinian fitness.
\newblock {\em Theoretical Population Biology}, 72(3):323 -- 345, 2007.

\end{thebibliography}
\end{document}